\documentclass[10pt,a4paper]{article}
\usepackage{amsmath}
\usepackage{graphicx,color}

\newcommand{\dP}[2]{\frac{\partial#1}{\partial#2}}

\newcommand{\hm}{h_\text{m}}

\newcommand{\Rl}{R_\text{l}}
\newcommand{\Rm}{R_\text{m}}
\newcommand{\rhos}{\rho_\text{s}}

\begin{document}

\title{Spreading of a liquid drop over a plane rigid substrate due to intermolecular forces}
\author{Alexander N. Tyatyushkin}

\maketitle

\begin{abstract}
Self-similar solutions of the equation that describes spreading of a liquid layer due to intermolecular forces are found. It is supposed that, when the thickness of the layer reaches some magnitude of the order of the molecular size, it turns into a monomolecular layer, which can be liquid or gaseous. To describe the spreading of a drop, the solutions of the equations that describe evolution of liquid or gaseous monomolecular layers are matched with the self-similar solutions with using relevant boundary conditions.
\end{abstract}

\section{Introduction}

Coating surfaces with monomolecular layers is an important part of many modern technologies. Usually it is preceded by spreading under the action of intermolecular forces. Self-similar solutions of the equation that describes plane spreading due to intermolecular forces represent profiles of drops without contact lines \cite{L.M.R.}.
In the present work, self-similar solutions are found also for axisymmetrical spreading. Both plane and axisymmetrical solutions are matched with the solutions of equations that describe plane or axisymmetrical spreading of monomolecular layers.

\section{General equation describing spreading}

\begin{figure}[h]
\centerline{
\includegraphics[width=\textwidth]{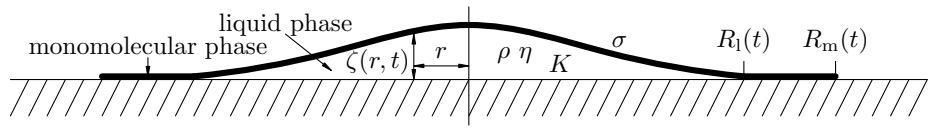}
}
\caption{Spreading drop (the monomolecular layer is in liquid two-dimensional state)}
\label{1}
\end{figure}
Consider a liquid layer spreading over a plane rigid substrate (see Fig.\ref{1}). The density $\rho$ and viscosity $\eta$ of the liquid are much larger than those of the environment (air). Let the layer be sufficiently thin and have sufficiently small maximal local inclination of the free surface so that the lubrication theory approximation can be used. The rigid substrate is implied to be much thicker than the characteristic length of intermolecular forces. Then axisymmetrical and plane spreading is described by the following equation for the local thickness of the liquid layer, $\zeta(r,t)$, 
\begin{equation}
\dP{\zeta}{t}
=
\frac{1}{3\eta}
\frac{1}{r^m}
\dP{}{r}
\left(
 r^m
 \zeta^3
 \left\{
  \rho g \dP{\zeta}{r}
  -
  \sigma
  \dP{}{r}
   \left[
    \frac{1}{r^m} \dP{}{r} \left(r^m \dP{r}{\zeta}\right)
   \right]
  +
  3K\zeta^{-4} \dP{\zeta}{r}
 \right\}
\right)
.
\label{m:se}
\end{equation}
Here, $m=0$ for plane symmetric spreading and $m=1$ for axisymmetrical spreading, $r$ is the distance from the plane ($m=0$) or axis ($m=1$) of symmetry, $g$ is the gravity acceleration, $\sigma$ is the surface tension, $K$ is the effective Hamaker constant (see \cite{L.M.R.}). If the term with $\sigma$ in (\ref{m:se}) is neglected, the resulting equation coincides with that written down in \cite{L.M.R.}.

Self-similar solutions of the equation (\ref{m:se}) exist only for special cases when one of the forces, gravitational, capillary or intermolecular, dominates so that the two others can be neglected. In what follows, the thickness of the layer is supposed to be so small that gravitational and capillary forces can be neglected.

\section{Self-similar spreading due to intermolecular forces}

The self-similar solutions of the equation that results from (\ref{m:se}) when the terms with $g$ and $\sigma$ being neglected are as follows
\begin{align*}
m=0
\!:\ 
\zeta(r,t)
&=
\dfrac
 {2H_0 R^2(t) \left(1+\dfrac{t}{T_0}\right)^{-1}}
 {2R^2(t)+\lambda r^2}
,\quad 
R(t)
=
R_0\left(1+\dfrac{t}{T_0}\right)
,\\ 
H_0
&=
\dfrac
 {\sqrt{\dfrac{\lambda}{2}}\dfrac{V}{R_0}}
 {\arctan\sqrt{\dfrac{\lambda}{2}}}
,\quad 
T_0
=
\dfrac
 {\dfrac{1}{2}\sqrt{\dfrac{1}{2\lambda}}\dfrac{V\eta}{K}}
 {\arctan\sqrt{\dfrac{\lambda}{2}}}
,
\end{align*}
\begin{align*}
m=1
\!:\ 
\zeta(r,t)
&=
\dfrac
 {2H_0 R^2(t) \exp\left(-\dfrac{2t}{T_0}\right)}
 {2R^2(t) + \lambda r^2}
,\quad 
R(t)=R_0\exp\dfrac{t}{T_0}
,\\ 
H_0
&=
\dfrac{\lambda}{2\pi\ln\left(1+\lambda/2\right)} \dfrac{V}{R_0^2}
,\quad 
T_0
=
\dfrac{1}{\pi\ln\left(1+\lambda/2\right)} \dfrac{V\eta}{K}
,
\end{align*}
where $\lambda$ is an arbitrary dimensionless parameter, $V$ is the volume (volume per unit transversal length for $m=0$) of the liquid contained in the region $r<R_0$ at the initial instant $t=0$, $R_0$ being an arbitrarily chosen distance. For $m=0$ the solutions coincide with those in \cite{L.M.R.}.

\section{Formation of a monomolecular layer}

When a layer of a liquid reaches some thickness, $\hm$, of the order of the molecular size, the liquid forms a monomolecular layer which can be in liquid or gaseous two-dimensional state. 
The moving boundary between monomolecular and liquid phases (see Fig.\ref{1}) is determined by some value, $\Rl(t)$, of the variable $r$ for which the following equation and initial condition are satisfied
\begin{equation}
\zeta\left(\Rl(t),t\right)
=
\hm
,\quad
\Rl(0)
=
R_0
.
\label{bc:zeta}
\end{equation}
Regarding that all mass of the liquid is in liquid phase at $t=0$ and using (\ref{bc:zeta}), one determines $\Rl(t)$ in terms of $R_0$ and $V=M/\rho$, where $M$ is the total mass of the liquid contained in the spreading drop: 
\begin{align*}
m&=0
\!:\  
\Rl(t)
=
\sqrt{\dfrac{2}{\lambda} \dfrac{T-t}{T_0} \dfrac{T_0+t}{T_0}} R_0
,\qquad 
T
=
\left(\dfrac{H_0}{\hm}-1\right)T_0
,\\
m&=1
\!:\ 
\Rl(t)
=
R_0
\sqrt{\dfrac{2}{\lambda}} 
\sqrt{\dfrac{H_0}{\hm}-\exp\dfrac{2t}{T_0}} 
,\qquad 
T
=
\dfrac{1}{1}T_0 \ln\left(\dfrac{H_0}{\hm}\right)
,
\end{align*}
where $\lambda$ is the non-zero positive solution of the equation
\begin{align*}
m&=0
\!:\ 
\sqrt{\dfrac{2}{\lambda}} \arctan\sqrt{\dfrac{\lambda}{2}}
=
\dfrac{V}{\hm R_0} \dfrac{2}{\lambda+2}
,\\
m&=1
\!:\ 
\ln\left(1+\lambda/2\right)
=
\dfrac{V}{\pi\hm R_0^2} \dfrac{\lambda}{\lambda+2}
.
\label{eq:lambda}
\end{align*}

\section{Plane or axisymmetrical spreading with formation of a liquid monomolecular layer}

The liquid monomolecular layer forms a distinct boundary with the uncovered region of the surface of the substrate which is determined by some value, $\Rm(t)$, of the variable $r$. Let the surface mass density of the layer, $\rhos$, be constant. Then, using the condition of the conservation of the total mass of the spreading liquid, one obtains
\begin{align*}
m&=0
\!:\  
\Rm(t)
=
\Rl(t)
+
\dfrac{\rho}{\rhos} 
\dfrac{V}{2} 
\left(
1 
-
\dfrac
 {\arctan\sqrt{\dfrac{T-t}{T_0+t}}}
 {\arctan\sqrt{\dfrac{\lambda}{2}}} 
\right)
,\\
m&=1
\!:\ 
\Rm(t)
=
\sqrt{
\Rl^2(t)
+
\dfrac{\rho}{\rhos} \dfrac{V}{\pi}
\left(
 1
 -
 \dfrac{2}{\ln\left(1+\lambda/2\right)} \dfrac{T-t}{T_0}
\right)
}
.
\end{align*}
At $t=T$ all the liquid is in monomolecular phase and stops spreading.

\section{Plane or axisymmetrical spreading with formation of a gaseous monomolecular layer}

\newcommand{\Ds}{D_\mathrm{s}}
If the monomolecular layer is gaseous, it has no boundary with the uncovered surface, and the distribution of the surface mass density, $\rhos(r,t)$, satisfies the differential equation for the surface diffusion 
\begin{equation*}
\dP{\rhos}{t}
=
\Ds
\frac{1}{r^m}
\dP{}{r}
\left(r^m\dP{\rhos}{r}\right)
,\quad
r>\Rl(t)
,
\end{equation*}
where $\Ds$ is the surface diffusion coefficient, with the boundary and initial conditions that follow from continuity of surface mass flux and mass densities on the boundary between the phases and absence of monomolecular phase at $t=0$
\begin{equation*}
-
\left.\Ds \dP{\rhos}{r}\right|_{r=\Rl(t)}
=
\dfrac{K\rho}{\eta} 
\dfrac
 {\dfrac{\lambda}{2}\dfrac{2\Rl(t)}{R^2(t)}}
 {1+\dfrac{\lambda}{2}\dfrac{\Rl^2(t)}{R^2(t)}}
,\qquad
\rhos(\Rl(t),t)=\rho\hm
,\qquad
\rhos(r,0)=0
.
\end{equation*}
At $t=T$ all the liquid is in monomolecular phase, but continues spreading so that
\begin{equation*}
\rhos(r,t)
\sim
\dfrac{\rho{V}}{2^{m+1} (\pi\Ds{t})^{(m+1)/2}}
\exp\left(-\dfrac{r^2}{4\Ds{t}}\right)
\text{ as }
t\to\infty
.
\end{equation*}

\section{Discussion}
The functions $\Rm(t)$ or $\rhos(r,t)$ for spreading with formation of liquid or gaseous monomolecular layers can be measured in experiments. In spite of the fact that the proposed models are based on strong simplifying assumptions, one may expect their satisfactory agreement at least with some experiments. In any case, the models allow introducing into consideration various additional physical processes to achieve satisfactory agreement with measured data.

Usually, when a drop begins to spread in an experiment, the conditions for existence of self-similar solutions are not satisfied. So, in an experimental verification of the formulas presented above, the initial instant $t_0$ should be introduced by the substitution $t{\to}t-t_0$ and regarded as some fitting parameter.

\end{document}